\documentclass[6pt,a4paper]{article}
\usepackage{bbm}
\usepackage{amsfonts}
\usepackage{mathrsfs}
\usepackage {amssymb}
\usepackage {amsmath}
\usepackage{amsthm}
\usepackage{latexsym}
\usepackage{booktabs}
\def\dse#1{\vskip 0.6cm\noindent
        {\large\bf #1}
        \vskip 0.4cm}
\def\dec#1{\vskip 0.6cm\noindent
        {\large\bf #1}
        \vskip 0.4cm}
\usepackage{lastpage}
\usepackage{epsfig}

\begin{document}

\begin{center}
\textbf{\large{Reversible Codes and Its Application to Reversible DNA Codes over $F_{4^k}$}}\footnote { E-mail addresses:
chenlei940511@126.com(L. Chen), lijin\_0102@126.com(J. Li), sunzhonghuas@163.com(Z. Sun).
This research supported by the National Natural Science Foundation of China(NO. 11501156) and the Anhui Provincial Natural Science Foundation (No. 1508085SQA198). }

\end{center}
\begin{center}
{ { Lei Chen \  Jin Li  \  Zhonghua Sun} }
\end{center}

\begin{center}
\textit{School of Mathematics, Hefei University of
Technology, Hefei 230009, Anhui, P.R.China %\\
 %$^{2}$National Mobile Communications Research Laboratory,
%Southeast University, Nanjing  210096,  P.R.China
}
\end{center}

\noindent\textbf{Abstract:} Coterm polynomials are introduced by Oztas et al. [a novel approach for constructing reversible codes and applications to DNA codes over the ring $F_2[u]/(u^{2k}-1)$, Finite Fields and Their Applications 46 (2017).pp. 217-234.], which generate reversible codes. In this paper, we generalize the coterm polynomials and construct some reversible codes which are optimal codes by using $m$-quasi-reciprocal polynomials. Moreover, we give a map from DNA $k$-bases to the elements of $F_{4^k}$, and  construct reversible DNA codes over $F_{4^k}$ by DNA-$m$-quasi-reciprocal polynomials.\\


\dec{1~~Introduction}
DNA is a molecule that carries most of the genetic instruction for the functions of cells. DNA sequences consist of Adenine ($A$), Guanine ($G$), Cytosine ($C$), Thymine ($T$) nucleotides. DNA has strands that are governed by the rule called Watson Crick complement (WCC). According to WCC, $A$ and $G$ are complement of $T$ and $C$, respectively. DNA computing started by the pioneer paper written by Leonard Adleman who solved a famous Hamiltonian path problem by using DNA molecules in [1]. Quaternary reversible complement cyclic  codes was used to generate DNA codes first introduced in [2], which considered the reverse constraint. Later, in [3], Gaborit and King proposed new constructions for
DNA codes satisfying either a reverse-complement constraint, a GC-content constraint, or both, that
were derived from additive and linear codes over four-letter alphabets, in particular on codes
over GF(4), they constructed new DNA codes that are in many cases better than previously known codes.
In [4], DNA
codes were generated by considering the four-element field GF(4) and construction of large sets of DNA codewords was achieved.
Then more cyclic DNA codes over the quaternary ring  $F_2+uF_2$ was  constructed in [5-7].\\

 In 2012, Yildiz and Siap first related DNA pairs with the ring $F_2[u]/(u^4-1)$  and  obtained some DNA codes by using the cyclic codes over $F_2[u]/(u^4-1)$ in [8].  In 2013,  Oztas et al. constructed reversible codes of odd length by lifted polynomials over the field $F_{16}$ and gave its applications to DNA codes in [9]. Byram et al. obtained some DNA codes over $F_4[v]/(v^2-v)$ that attain the Griesmer bound in [10]. Zhu et al. studied the structure of cyclic DNA codes of arbitrary lengths over the ring $F_2+uF_2+vF_2+uvF_2$ in [11]. In  [12], Limbachiya et al.  established an one to one correspondence between the elements of
 the ring $Z_4+wZ_4$ and all the DNA codewords of length 2, and gave several new classes of the DNA codes which satisfied reverse and reverse-complement constraints.    Dinh et al. considered the structure of cyclic DNA codes of odd length over the finite commutative ring $F_2+uF_2+vF_2+uvF_2+v^2F_2+uv^2F_2$ in [13]. \\

  Recently, Oztas et al. generalized the $4^k$-lifted polynomials which led to reversible and reversible complement DNA codes over $F_{4^{2k}}$ in [14]. Furthermore, they identified k-bases of DNA with elements in the ring $F_2[u]/(u^{2k}-1)$, and solved the reversibility and complement problems in DNA codes over this ring by using a form of coterm polynomials in [15]. They found reversible and reversible-complement codes that were not necessarily linear cyclic codes. Motivate by this, we generalize the coterm polynomials  and use so-called $m$-quasi-reciprocal polynomial ¡± to construct reversible codes.\\

  The rest of this paper is organized as follows.
  In Section 2 we recall basic notions for DNA codes.
  In Section 3, we define the $m$-quasi-reciprocal polynomials, then we construct some reversible codes over $F_q$ which are optimal codes by using  $m$-quasi-reciprocal polynomials. In Section 4, we find that there exist a map between DNA $k$-bases and the elements in $F_{4^{k}}$, and we use DNA-$m$-quasi-reciprocal polynomials to obtain reversible DNA codes over $F_{4^{k}}$. Section 5 concludes the paper.

\dec{2~~Preliminaries}

Let $F_q$ be a finite field of $q$ elements, and $n$ be a positive integer. A linear code $\mathcal{C}$ over $F_q$ of length $n$ is a $F_q$-linear subspace of $F_q^n$. For any $(a_0,a_1,\ldots,a_{n-1})\in F_q^n$, we give a permutation $\pi$,
\begin{center}
	$\pi(a_0,a_1,\ldots,a_{n-1})=(a_{n-1},a_0,a_1,\ldots,a_{n-2})$.
\end{center}
 $\pi$ is called a \emph{cyclic right shift}. The permutation ${\pi}^{-1}$ denotes \emph{cyclic left shift},
$$\pi^{-1}(a_0,a_1,\ldots,a_{n-1})=(a_1,a_2,\ldots,a_{n-1},a_0).$$

For any positive integer $i$, $\pi^{i}(a_0,a_1,\ldots,a_{n-1})=( a_{n-i} ,$ $ a_{n-i+1} , \ldots , a_{n-1} , a_0 , a_1 , \ldots , a_{n-i-1})$
is called a cyclic right shift $i$ times, $\pi^{-i}(a_0,a_1,\ldots,a_{n-1})=(a_i,$ $a_{i+1},\ldots,a_{n-1},a_0,a_1,\ldots,a_{i-1})$
is called a cyclic left shift $i$ times.
It is easy to see, $\pi^0(a_0,$ $a_1,\ldots,a_{n-1})=(a_0,a_1,\ldots,a_{n-1})$, $\pi^n(a_0,a_1,\ldots,a_{n-1})=\pi^{-n}(a_0,a_1,\ldots,a_{n-1})$.

For any $\textbf{c}=(c_0,c_1,\ldots,c_{n-1})\in F_q^n$, we let $\textbf{c}^r=(c_{n-1},c_{n-2},\ldots,c_1,c_0)$ denote the \emph{reverse} of $\textbf{c}$ and $\textbf{c}$ is self-reversible if $\textbf{c}=\textbf{c}^r$. Naturally, for $\textbf{a}=(a_0,a_1,\cdots,a_{l-1}) \in F_q^l$, $1\leq l\leq n$, we also let $\textbf{a}^r=(a_{l-1},a_{l-2},\ldots,a_1,a_0)$ and say $\textbf{a}$ is self-reversible if $\textbf{a}=\textbf{a}^r$. A linear code $\mathcal{C}$ is said to be \emph{reversible} if $\textbf{c}^r\in \mathcal{C}$ for all $\textbf{c}\in \mathcal{C}$.\\

 Next we will give some basic background on DNA codes. As for the given DNA codeword $\textbf{x}=(x_1 x_2 \cdots x_{n-1} x_n)$, where $x_i \in D=\{A,T,C,G\}$, for ${i=1,2,\cdots,n}$ and $A^c=T,T^c=A,C^c=G,G^c=C$ according to WCC. The reverse of DNA codeword is defined as $\textbf{x}^r=(x_n x_{n-1} \cdots x_2 x_1)$, the complement of DNA codeword is defined as $\textbf{x}^c=(x_1^c x_2^c \cdots x_{n-1}^c x_n^c)$, the reverse complement of DNA codeword is defined as $\textbf{x}^{rc}=(x_n^c x_{n-1}^c \cdots x_2^c x_1^c)$.\\

 The minimum Hamming distance $d$ of linear $\mathcal{C}$ is defined as $\min \{d(\textbf{a},\textbf{b})|\textbf{a}\neq \textbf{b} ,\forall \textbf{a},\textbf{b}\in \mathcal{C}\}$.

 A DNA code $\mathcal{C}$ of length $n$ is defined as a set of codewords $(x_1x_2\ldots x_n)$ where $x_i\in D=\{ A, T, C,G \}$, and $\mathcal{C}$ satisfies some or all of the following constraints:

\begin{itemize}
	\item [(i)] The Hamming distance constraint\\
If $d(\textbf{x},\textbf{y})\geq d$ for all $\textbf{x},\textbf{y}\in \mathcal{C}$, with $\textbf{x}\neq \textbf{y}$, for some prescribed minimum distance $d$.
	\item [(ii)] The reverse constraint\\
If $d(\textbf{x}^r,\textbf{y})\geq d$ for all $\textbf{x},\textbf{y}\in \mathcal{C}$, including $\textbf{x}=\textbf{y}$.
	\item [(iii)] The reverse-complement constraint\\
If $d(\textbf{x}^{rc},\textbf{y})\geq d$ for all $\textbf{x},\textbf{y}\in \mathcal{C}$, including $\textbf{x}=\textbf{y}$.
	\item [(iv)] The fixed $GC$-content constraint\\
If any codeword contains the same number of $G$ and $C$.
\end{itemize}

The purpose of the first three constraints is to reduce the probability of non-specific hybridization. The fixed $GC$-content constraint is used to obtain similar melting temperatures. In this paper we mainly study the reversible DNA codes which satisfy the reverse constraint and in the end of the paper we also study the reversible complement DNA codes which satisfies the reverse-complement constraint.

\dec{3~~Reversible codes over $F_q$}

We first give the definition of $m$-quasi-reciprocal polynomial. Next we give some reversible codes by using $m$-quasi-reciprocal polynomials to construct reversible codes over different fields which have the best possible parameters according to codetables in [16] or attaining the Griesmer bound in [18].\\

 \textbf{Lemma 3.1$^{[18]}$}(Griesmer bound) For a linear [n,k,d] code $\mathcal {C}$ over a finite field $F_q$, that is, a $k$-dimensional subspace of a vector space $F_q^n$ with minimum Hamming distance $d$, the Griesmer bound is $n\geq \sum_{i=0}^{k-1}\lceil d/q^i \rceil$.\\

In the following, we let $m$ and $t$ be any nonnegative integers and $n$ be any positive integer.\\

\textbf{Definition 3.2} $f(x)=a_0+a_1x+\cdots+a_{n-1}x^{n-1}\in F_q[x]$ is called a $m$-quasi-reciprocal polynomial, if $a_i=a_{m-1-i}$ for $0\leq i\leq\lfloor{(m-1)/2}\rfloor$ and $a_j=a_{m+n-1-j}$ for $m\leq j\leq\lfloor{{(m+n-1)}/2}\rfloor$. In particular, if $m=0$, $m$-quasi-reciprocal polynomial is self-reciprocal polynomial. \\

Throughout this paper, $F_4=\{ 0, \alpha, 1,\alpha+1\}$, $\alpha$ is the root of irreducible polynomial $f(x)=x^2+x+1$, i.e., $\alpha$ is a primitive element of $F_4$.

For example $f(x)=\alpha+ \alpha x+\alpha^2 x^2+x^3+\alpha ^2x^4$ is a $2$-quasi-reciprocal polynomial over $F_4$. \\

Let $\varnothing \neq S \subseteq F_q^n$, $\langle S\rangle$ denote the $F_q$-linear subspace of $F_q^n$ which generated by the set $S$. The goal in this paper is to construct reversible codes by choosing special set $S$. \\

\textbf{Lemma 3.3 $^{[9]}$ }If $\forall \textbf{s} \in S$, we have $\textbf{s}^r \in S$. Then $\langle S \rangle$ is a reversible code.\\

For any $\textbf{c}=(c_0,c_1,\ldots,c_{n-1})\in F_q^n$, polynomial representation of $\textbf{c}$ is $c(x)=c_0+c_1x+\cdots+c_{n-1}x^{n-1}$. Next, each $\textbf{c} \in F_q^n$ is identified with its polynomial representation $c(x)$.\\

\textbf{Theorem 3.4} If $c(x)$ is a $m$-quasi-reciprocal polynomial. Then $\langle S_t\rangle$ is a reversible code of length $n$ over $F_q$, where  $$S_t={\{\pi^t(\textbf{c}),\pi^{t-1}(\textbf{c}),\cdots,\pi^1(\textbf{c}),\pi^0(\textbf{c}),\pi^{-m}(\textbf{c}),\pi^{-m-1}(\textbf{c}),\cdots,\pi^{-m-t+1}(\textbf{c}),\pi^{-m-t}(\textbf{c})\}}.$$

Moreover,
\begin{itemize}
	\item [(i)] if $m=0$ and $n$ is even, then $\langle E_t \rangle$ is a reversible code, where $E_t=\{S_t, \pi^{n/2}(\textbf{c})\}$.
	\item [(ii)] if $1\leq m\leq n-2$ and $m$ is even, then $\langle E_t^{(1)} \rangle$ is a reversible code, where $E_t^{(1)}=\{S_t, \pi^{-m/2}(\textbf{c})\}$.
	\item [(iii)] if $1\leq m\leq n-2$ and $n-m$ is even, then $\langle E_t^{(2)} \rangle$ is a reversible code, where $E_t^{(2)}=\{S_t, \pi^{(n-m)/2}(\textbf{c})\}$.
	\item [(iv)] if $1\leq m\leq n-2$, $m$ is even and $n-m$ is even, then $\langle E_t^{(3)} \rangle$ is a reversible code, where $E_t^{(3)}=\{S_t, \pi^{-m/2}(\textbf{c}), \pi^{(n-m)/2}(\textbf{c})\}$.
\end{itemize}

\textbf{Proof.} Since $c(x)$ is a $m$-quasi-reciprocal polynomial,  $(c_0,c_1,\ldots,c_{m-1})$, $(c_m,c_{m+1},\ldots,c_{n-1})$ are self-reversible.
For any positive integer $i$, we have
\begin{center}
	${(\pi^i(\textbf{c}))}^r=\pi^{-i-m}(\textbf{c})$.
\end{center}
If $i=0$, ${(\pi^0(\textbf{c}))}^r=\pi^{-m}(\textbf{c})$, then by applying induction on $i$, we can get the above equation. By Lemma 3.3, $S_t$ can generate reversible codes.\\
Note that
\begin{itemize}
	\item [(i)] when $m=0$ and $n$ is even, we can get
		$\pi^{n/2}(\textbf{c})=({\pi^{n/2}(\textbf{c})})^r$.
	\item [(ii)] when $m$ is even, we can get $({\pi^{-{m/2}}(\textbf{c})})^r={\pi^{-{m/2}}(\textbf{c})}$.
	\item [(iii)] when $n-m$ is even, we can get $(\pi^{{(n-m)/2}}(\textbf{c}))^r=\pi^{{(n-m)/2}}(\textbf{c})$.
	\item [(iv)] when $m$ is even and $n-m$ is even, we can get $({\pi^{-{m/2}}(\textbf{c})})^r={\pi^{-{m/2}}(\textbf{c})}$,  $(\pi^{{(n-m)/2}}(\textbf{c}))^r=\pi^{{(n-m)/2}}(\textbf{c})$.
\end{itemize}

Then, by Lemma 3.3, $E_t$, $E_t^{(1)}$, $E_t^{(2)}$, $E_t^{(3)}$ can generate reversible codes in the conditions of (i),(ii),(iii),(iv).
\\

\textbf{Remark.} The dimension of reversible code $\langle S_t\rangle$, $\langle E_t\rangle$, $\langle E_t^{(1)}\rangle$, $\langle E_t^{(2)}\rangle$ or $\langle E_t^{(3)}\rangle$  in Theorem 3.4 is not necessary equal to the number of elements in the set $ S_t$, $E_t$, $E_t^{(1)}$, $E_t^{(2)}$ or $E_t^{(3)}$. For example, let $\textbf{c}=(a,a,b,c,b)\in F_q^5$. Clearly, $a+ax+bx^2+cx^3+bx^4$ is a $2$-quasi-reciprocal polynomial. Let
\begin{align*}
S_1&=\{\pi^1(\textbf{c}),\pi^0(\textbf{c}),\pi^{-2}(\textbf{c}),\pi^{-3}(\textbf{c})\},\\
S_2&=\{\pi^2(\textbf{c}),\pi^1(\textbf{c}),\pi^0(\textbf{c}),\pi^{-2}(\textbf{c}),\pi^{-3}(\textbf{c}),\pi^{-4}(\textbf{c})\}.
\end{align*}
The reversible codes $\langle S_1\rangle$ and $\langle S_2 \rangle$ have the same dimension, because of $(\pi^1(\textbf{c}))^r=\pi^{-3}(\textbf{c})=\pi^{2}(\textbf{c})$, $(\pi^2(\textbf{c}))^r=\pi^{-4}(\textbf{c})=\pi^1(\textbf{c})$, then $\langle S_1\rangle$ = $\langle S_2 \rangle$ .\\

Throughout this paper, $f(x)=x^3+x^2+x+\alpha$ is an irreducible polynomial over $F_4$ and we let $\gamma$ be the root of $f(x)$ in $F_{64}$, i.e., $\gamma$ is a primitive element of $F_{64}$.\\

\textbf{Example 3.5}  Let $\textbf{c}=(\gamma^2,\gamma^2,1,1,\gamma,0,\gamma,1,1)\in F_{64}^9$. It is easy to see $c(x)$ is a $2$-quasi-reciprocal polynomial over $F_{64}$.

\begin{itemize}
	\item [(i)] If $t=1$, $\mathcal{C}_1=\langle{S_1}\rangle$ is a $[9,4,6]$ reversible code over $F_{64}$, where $S_1={{\{}\pi^1(\textbf{c}),}\pi^0(\textbf{c}),$ $\pi^{-2}(\textbf{c}),\pi^{-3}(\textbf{c}){\}}$.
	The generator matrix of $\mathcal{C}_1$ is
	\begin{center}
		$\textbf{G}_1=\left(
		\begin{array}{ccccccccc}
		{\gamma}^2 & {\gamma}^2 & 1 & 1 & {\gamma} & 0 & \gamma & 1 & 1 \\
		1 & 1 & {\gamma} & 0 & \gamma & 1 & 1 & {\gamma}^2 & {\gamma}^2 \\
        1 & {\gamma}^2 & {\gamma}^2 & 1 & 1 & {\gamma} & 0 & \gamma & 1 \\
        1 & {\gamma} & 0 & \gamma & 1 & 1 & {\gamma}^2 & {\gamma}^2 & 1 \\
		\end{array}
		\right).
		$
	\end{center}
	
     \item [(ii)] If $t=1$, $\mathcal{C}_2=\langle{E_1^{(1)}}\rangle$ is a $[9,5,4]$ reversible code over $F_{64}$, where $E_1^{(1)}={{\{}\pi^1(\textbf{c}),}\pi^0(\textbf{c}),$ $\pi^{-2}(\textbf{c}),\pi^{-3}(\textbf{c}), \pi^{-1}(\textbf{c}){\}}$.
	The generator matrix of $\mathcal{C}_2$ is
	\begin{center}
		$\textbf{G}_2=\left(
		\begin{array}{ccccccccc}
		{\gamma}^2 & {\gamma}^2 & 1 & 1 & {\gamma} & 0 & \gamma & 1 & 1 \\
		1 & 1 & {\gamma} & 0 & \gamma & 1 & 1 & {\gamma}^2 & {\gamma}^2 \\
        1 & {\gamma}^2 & {\gamma}^2 & 1 & 1 & {\gamma} & 0 & \gamma & 1 \\
        1 & {\gamma} & 0 & \gamma & 1 & 1 & {\gamma}^2 & {\gamma}^2 & 1 \\
        {\gamma}^2 & 1 & 1 & {\gamma} & 0 & \gamma & 1 & 1 & {\gamma}^2 \\
		\end{array}
		\right).
		$
	\end{center}
\end{itemize}

We note that $\mathcal{C}_1$ is a $[9,4,6]$ reversible code over $F_{64}$, and $\sum_{i=0}^{k-1}\lceil d/q^i\rceil=\sum_{i=0}^3\lceil 6/{64}^i\rceil=9$, then $\mathcal{C}_1$ is a MDS code and attains the Griesmer bound. $\mathcal{C}_2$ is a $[9,5,4]$ reversible code over $F_{64}$, then $\mathcal{C}_2$ is a almost MDS code.\\

 We give more examples of optimal codes in Table 1 with added property of being reversible.

\begin{table}[htbp]
	
	\caption{Some examples of reversible codes generated via Theorem $3.4$ that are optimal codes.}
	\begin{tabular}{cccccc}
		\toprule
		$n$ & $q$ & $m$ & Coefficients of the $m$-quasi-reciprocal polynomial & codes & parameters  \\
		\midrule
		$11$ & $4$ & $2$ & $(1,1,0,{\omega}^2,{\omega},1,0,1,{\omega},{\omega}^2,0)$ & $\langle S_0\rangle$ & $[11,2,8]$  \\
		$11$  & $4$ & $3$ & $(1,{\omega}^2,1,{\omega}^2,{\omega}^2,{\omega},1,1,{\omega},{\omega}^2,{\omega}^2)$ & $\langle E_0^{(2)}\rangle$ & $[11,3,7]$  \\
		$13$  & $4$ & $3$ & $(1,0,1,{\omega}^2,{\omega},{\omega}^2,{\omega},1,1,{\omega},{\omega}^2,{\omega},{\omega}^2)$ & $\langle S_2\rangle$ & $[13,6,6]$ \\
		$5$  & $4$ & $0$ & $({\omega}^2,{\omega},0,{\omega},{\omega}^2)$ & $\langle S_1\rangle$ & $[5,2,4]^*$ \\
        $13$  & $3$ & $4$ & $(0,2,2,0,1,0,1,2,1,2,1,0,1)$ & $\langle S_0\rangle$ & $[13,2,9]$ \\
		$11$  & $8$ & $3$ & $(1,1,1,{\omega}^4,{\omega}^2,{\omega}^5,{\omega}^3,{\omega}^3,{\omega}^5,{\omega}^2,{\omega}^4)$ & $\langle S_0\rangle$ & $[11,2,9]^{**}$ \\
		$21$  & $9$ & $5$ & $({\omega}^3,{\omega},0,{\omega},{\omega}^3,{\omega}^2,{\omega}^3,{\omega},0,1,1,1,1,1,1,1,1,0,{\omega},{\omega}^3,{\omega}^2)$ & $\langle S_0\rangle$ & $[21,2,18]$ \\
		$7$  & ${25}$ & $3$ & $({\omega},0,{\omega},{\omega}^8,1,1,{\omega}^8)$ & $\langle S_0\rangle$ & $[7,2,6]^*$ \\
		$7$ & ${25}$ & $3$ & $({\omega},1,{\omega},{\omega}^3,0,0,{\omega}^3)$ & $\langle E_0^{(2)}\rangle$ & $[7,3,5]^*$ \\
		$9$  & ${27}$ & $1$ & $(1,{\omega},{\omega}^3,1,0,0,1,{\omega}^3,{\omega})$ & $\langle E_1^{(2)}\rangle$ & $[9,5,4]^{**}$ \\
		$6$  & ${64}$ & $0$ & ${(\omega}^2,\omega,0,0,\omega,{\omega}^2)$ & $\langle S_1\rangle$ & $[6,3,4]^*$ \\

		\bottomrule
	\end{tabular}
	
	$^*$: MDS code, $^{**}$: almost MDS code, $\omega$: primitive element of corresponding fields.
\end{table}

\dec{4~~Reversible DNA codes over $F_{4^{k}}$}

As for constructing reversible DNA codes over $F_{4^k}$, the most difficult and interesting problem is to provide an optimal matching between DNA $k$-bases and the elements in $F_{4^k}$. For $k=1$, in [4], Abulraub and Ghrayeb gave the matching between the DNA alphabets and the elements in $F_4$. For $k=2l$, Oztas and Siap introduced the matching process between DNA $2l$-bases and the elements of  $F_{4^{2l}}$ under some circumstances in [14]. However, as for general $k$, reversible DNA codes over $F_{4^{k}}$ haven't be resolved. In this section we first give a map $\overline{\zeta}$ between DNA $k$-bases and the elements in $F_{4^k}$. Next, we  use DNA-$m$-quasi-reciprocal polynomial to construct reversible DNA codes over $F_{4^{k}}$. From now on, we make the following assumptions for the rest of this paper to match DNA $k$-bases with the elements in $F_{4^{k}}$.

\begin{itemize}
	\item [(i)] Let $\omega$ be a primitive element of $F_{4^k}$ such that $\alpha=\omega^{\frac{4^k-1}3}$, where $\alpha$ is a primitive element of $F_4$.
	\item [(ii)] Let $m$ be a positive integer such that $\omega^m=1+\omega+\cdots+\omega^{k-1}$. We can obtain that $\{\omega,{\omega}^2,\ldots,\omega^{k-1},\omega^m \}$ is a basis of $F_{4^k}$ over $F_4$.
	\item [(iii)] $\zeta$ be the map from $D=\{ A, T, G, C\}$ to $F_4$ such that $\zeta(A)=0$, $\zeta(T)=1$, $\zeta(C)=\alpha$ and $\zeta(G)=1+\alpha$.
\end{itemize}

The element of $D^k$ is called a DNA $k$-base. In order to construct DNA codes, we establish a correspondence between DNA $k$-bases and the elements in $F_{4^k}$.\\

\textbf{Definition 4.1} Let $n,k$ be positive integers and $\beta_i=b_{i,1}b_{i,2}\ldots b_{i,k}$ be a DNA $k$-base, where $b_{i,j}\in{D}$, $1\leq i \leq n$ and $1\leq j \leq k$. Suppose $\beta_i^r=b_{i,k}b_{i,k-1}\ldots b_{i,2}b_{i,1}$, $\beta_i^c=b_{i,1}^cb_{i,2}^c\ldots b_{i,k}^c$. Let $B=(\beta_1,\beta_2,\ldots,\beta_{n})$,  $\overline{\zeta}$ be an one to one map from $D^k$ to $F_{4^{k}}$,
\begin{equation*}
\overline{\zeta}(\beta_i)=\omega \zeta(b_{i,1})+\omega^2 \zeta(b_{i,2})+\cdots+\omega^{k-1} \zeta(b_{i,k-1})+\omega^m\zeta(b_{i,k}),
\end{equation*}
Naturally, we generate the map $\overline{\zeta}$ to be an one to one map from $D^{kn}$ to $F_{4^{k}}^n$,
\begin{equation*}
\overline{\zeta}(\beta_1,\beta_2,\ldots,\beta_{n})=(\overline{\zeta}(\beta_1),\overline{\zeta}(\beta_2),\ldots,\overline{\zeta}(\beta_{n}))\in F^n_{4^{k}}.
\end{equation*}

\textbf{Remark.} It is easy to check that
 $\zeta(x^c)=\zeta(x)+1$, for $x\in D=\{A,T,C,G\}$.\\
Notice that $
\overline{\zeta}(\beta_i^c)=\omega \zeta(b_{i,1}^c)+\omega^2 \zeta(b_{i,2}^c)+\cdots+\omega^{k-1} \zeta(b_{i,k-1}^c)+\omega^m\zeta(b_{i,k}^c)=\omega (\zeta(b_{i,1})+1)+\omega^2 (\zeta(b_{i,2})+1)+\cdots+\omega^{k-1} (\zeta(b_{i,k-1})+1)+\omega^m(\zeta(b_{i,k})+1)=\omega \zeta(b_{i,1})+\omega^2 \zeta(b_{i,2})+\cdots+\omega^{k-1} \zeta(b_{i,k-1})+\omega^m\zeta(b_{i,k})+1=\overline{\zeta}(\beta_i)+1.\\
$
Then $\overline{\zeta}(x^c)=\overline{\zeta}(x)+1$, for $x\in D^k$.
Then we choose $\{\omega,{\omega}^2,\ldots,\omega^{k-1},\omega^m\}$ as a basis of $F_{4^k}$ over $F_4$.\\

Recall that, $\alpha$ is a primitive element of $F_4$, $\gamma$ is a primitive element of $F_{64}$ and $\gamma^3+\gamma^2+\gamma+{\alpha}=0$.\\

\textbf{Example 4.2} In $F_{64}$, it is now straightforward to prove that $1+\gamma+{\gamma}^2={\gamma}^{20}$ and $\alpha={\gamma}^{21}$.
 Considering  $AGT$ as a DNA 3-base, we can get the corresponding element in $F_{64}$ of AGT is
$${\zeta}(A){\gamma}+{\zeta}(G){\gamma}^2+{\zeta}(T){\gamma}^{20}={\gamma}^2 (1+{\alpha})+{\gamma}^{20}=1+{\gamma}+{\alpha}{\gamma}^2.$$

Next we will give an one to one map $\varphi$ from $F_{4^k}$ to $F_{4^k}$, which will give the reverse of the corresponding DNA $k$-base of the element in the field $F_{4^k}$.\\

\textbf{Lemma 4.3} For any $z=\omega z_1+\omega^2z_2+\cdots+\omega^{k-1}z_{k-1}+\omega^mz_k \in F_{4^k}$, $z_i{\in}F_4$, We denote $\varphi(z)=\omega z_{k}+\omega^2z_{k-1}+\cdots+\omega^{k-1}z_2+\omega^mz_1$. Then $\overline{\zeta}^{-1}(\varphi(z))$ gives the reverse of $\overline{\zeta}^{-1}(z)$, i.e., $({\overline{\zeta}}^{-1}(z))^r=\overline{\zeta}^{-1}(\varphi(z))$, where $\overline{\zeta}^{-1}$ is the inverse map of $\overline{\zeta}$. \\

\textbf{Proof.} On the one hand,
\begin{equation*}
\overline{\zeta}^{-1}(z)=\zeta^{-1}(z_1)\zeta^{-1}(z_2)\ldots\zeta^{-1}(z_{k-1})\zeta^{-1}(z_k).
\end{equation*}
Hence, the reverse of $\overline{\zeta}^{-1}(z)$ is $\zeta^{-1}(z_k)\zeta^{-1}(z_{k-1})\ldots\zeta^{-1}(z_2)\zeta^{-1}(z_1)$.

On the other hand, $\varphi(z)=\omega z_k +\omega^2z_{k-1}+\cdots+\omega^{k-1} z_2+\omega^m z_1$, the result then follows.\\

\textbf{Example 4.4} As for Example 4.2, we know that $\overline{\zeta}(AGT)=\gamma^2 (1+\alpha)+\gamma^{20}$ and $\overline{\zeta}(TGA)=\gamma+ \gamma^2 (1+{\alpha})=\varphi(\gamma^2 (1+\alpha)+\gamma^{20})$.\\

Corresponding to the Definition 3.2, we introduce the DNA-$m$-quasi-reciprocal polynomial.\\

\textbf{Definition 4.5} $\emph{B}^{\emph{DNA}}_m=({\beta}_0,{\beta}_1,\cdots,{\beta}_{n-1})$ is called $m$-quasi-reversible $n$-tuple of DNA $k$-bases, if ${\beta}_i={\beta}_{m-1-i}^r$ for $0 \leq i \leq \lfloor{(m-1)/2}\rfloor$ and ${\beta}_j={\beta}_{m+n-1-j}^r$ for $m\leq j\leq\lfloor{{(m+n-1)}/2}\rfloor$.
In particular, if $m=0$, $m$-quasi-reversible $n$-tuple of DNA $k$-bases is self-reciprocal $n$-tuple of DNA $k$-bases.

The polynomial representation of $\emph{B}^{\emph{DNA}}_m$, $f(x)=\overline{\zeta}(\beta_0)+\overline{\zeta}(\beta_1)x+\cdots+\overline{\zeta}(\beta_{n-1})x^{n-1} \in F_{4^k}[x]$, is called DNA-$m$-quasi-reciprocal polynomial. \\

\textbf{Example 4.6} $\emph{B}^{{DNA}}_2=(TCTC,CTCT,ATGC,GCTA,ATCG,CGTA)$ is a $2$-quasi-reversible $6$-tuple of DNA $4$-bases.\\

We consider an one to one map $\eta$ from $F_{4^k}$ to $F_{4^k}$, for any $a=\omega z_1+\omega^2z_2+\cdots+\omega^{k-1}z_{k-1}+\omega^mz_k\in F_{4^k}$, we have $\eta(a)=\omega z_2+\omega^2z_3+\cdots+\omega^{k-1}z_{k}+\omega^mz_1{\in}F_{4^k}$. It is easy to check that (i) $\eta^2(a)=\omega z_3+\omega^2z_4+\cdots+\omega^{k-2}z_{k}+\omega^{k-1}z_{1}+\omega^mz_2$, (ii) $\eta^{l}(a)=\eta(\eta^{l-1}(a))$ for $l\geq 2$, (iii) $\eta^{k}(a)=a$.

Naturally, we generate the map $\eta$ to be an one to one map from $F_{4^k}^n$ to $F_{4^k}^n$,
$${\eta}(c_1,c_2,\ldots,c_n)=({\eta}(c_1),{\eta}(c_2),\ldots,{\eta}(c_{n})) \in F^n_{4^{k}},$$
where $ c_i \in F_{4^k},1 \leq i\leq n$.\\

\textbf{Example 4.7} Let $b_1b_2\cdots b_k\in D^k$, if $a\in F_{4^k}$ and $\overline{\zeta}(b_1b_2\cdots b_k)=a$, then $\overline{\zeta}^{-1}(\eta(a))=b_2b_3\cdots b_kb_1$.\\

Next we use DNA-$m$-quasi-reciprocal polynomial to construct reversible DNA codes over $F_{4^k}$.\\

\textbf{Lemma 4.8} Let $c(x)=c_0+c_1x+\cdots+c_{n-1}x^{n-1}\in {F}_{4^k}[x]$ be a DNA-$m$-quasi-reciprocal polynomial, where $c_i=\omega a_{i,1}+\omega^2 a_{i,2}+\cdots+\omega^{k-1} a_{i,k-1}+\omega^m a_{i,k}$, $a_{i,j}\in F_{4}$, $0\leq i\leq n-1, 1\leq j\leq k $. Then
\begin{equation*}
(\overline{\zeta}^{-1}\eta^i(\pi^j(\textbf{c})))^r=\overline{\zeta}^{-1}(\eta^{k-i}(\pi^{-m-j}(\textbf{c})))
\end{equation*}
for $0\leq i\leq k-1$ and $0\leq j\leq n-1$. \\

\textbf{Proof.} On the one hand, $\pi^{-m-j}(\textbf{c})=(c_{m+j},c_{m+j+1},\ldots, c_{n-1},c_0,c_1,\ldots, c_{m+j-1})$. It follows from $c(x)$ is a DNA-$m$-quasi-reciprocal polynomial that
\begin{equation*}
\pi^{-m-j}(\textbf{c})=(\varphi(c_{n-j-1}),\varphi(c_{n-j-2}),\ldots, \varphi(c_{0}), \varphi(c_{n-1}),\varphi(c_{n-2}),\ldots, \varphi(c_{n-j})),
\end{equation*}

$\eta^{k-i}(\pi^{-m-j}(\textbf{c}))=(\eta^{k-i}(\varphi(c_{n-j-1})),\eta^{k-i}(\varphi(c_{n-j-2}))\ldots,
\eta^{k-i}(\varphi(c_{0})), \eta^{k-i}(\varphi(c_{n-1})), $\\

$\eta^{k-i}(\varphi(c_{n-2})) ,\ldots, \eta^{k-i}(\varphi(c_{n-j})))$.\\

On the other hand,\\
 $$\eta^i(\pi^j(\textbf{c}))=(\eta^{i}(c_{n-j}), \eta^{i}(c_{n-j+1}),\cdots, \eta^{i}(c_{n-1}), \eta^{i}(c_0), \eta^{i}(c_1) ,\cdots,\eta^{i}(c_{n-j-1}))$$

 For any $ 0\leq l\leq {n-1}$, we can easily find that $(\eta^{i}(c_l))^r=\eta^{k-i} (\varphi ((c_l))) $.\\
Hence, $\eta^{k-i}(\pi^{-m-j}(\textbf{c}))=(\eta^i(\pi^j(\textbf{c})))^r$. According to Lemma 4.3, the result then follows.\\

In order to construct reversible DNA codes over $F_{4^k}$, we introduce the map-code.\\

\textbf{Definition 4.9} Let $ \mathcal{C}$ be a reversible code, a map-code is defined as
$$ C_\eta ={\{} \sum_i \sum_{j=0}^{k-1}a_{ij} {\eta}^j (\textbf{p}_i):a_{ij} \in F_4, \textbf{p}_i \in\mathcal{C} {\}}$$
$C_{\eta}$ is a $F_4$-linear code.\\

\textbf{Theorem 4.10} If $c(x)$ is a DNA-$m$-quasi-reciprocal polynomial over $F_{4^{k}}$ and we can find that $C_{\eta}=\langle S_t \rangle_ \eta$, $C_{\eta}=\langle E_t \rangle_ \eta$, $C_{\eta}=\langle E_t^{(1)} \rangle_ \eta$, $C_{\eta}=\langle E_t^{(2)} \rangle_ \eta$ or $C_{\eta}=\langle E_t^{(3)} \rangle_ \eta$ is a map-code, where the set $S_t$, $E_t$, $E_t^{(1)}$, $E_t^{(2)}$ or $E_t^{(3)}$ comes from theorem 3.4. Then $\overline{\zeta}^{-1}(C_{\eta})$ is a reversible DNA code over $F_{4^k}$.\\

\textbf{Proof:} Here we only prove $\overline{\zeta}^{-1}(C_{\eta})$ is a reversible DNA code over $F_{4^k}$, where $C_{\eta}=\langle E_t^{(3)} \rangle_ \eta$.

For the set of $E_t^{(3)}={\{S_t,\pi^{-{m/2}}(\textbf{c}),\pi^{{(n-m)/2}}(\textbf{c})\}}$. Let $\textbf{g}$ be a typical codeword in $C_{\eta} $, then $\textbf{g}$ can be written as follows
$$\sum_j (\sum_{i=0}^{k-1} a_{j,i} \eta^i (\pi^j (\textbf{c}))) +\sum_{i=0}^{k-1} a_i\eta^i(\pi^{-m/2}(\textbf{c}))+\sum_{i=0}^{k-1} b_i\eta^i(\pi^{(n-m)/2}(\textbf{c}))$$
where $a_{j,i}, a_i,b_i\in F_4$. By Lemma 4.8,
$$
(\overline{\zeta}^{-1}(\textbf{g}))^r=\overline{\zeta}^{-1}(\sum_j (\sum_{i=0}^{k-1} a_{j,i} \eta^{k-i} (\pi^{-m-j} (\textbf{c})))+\sum_{i=0}^{k-1} a_i\eta^{k-i}(\pi^{-m/2}(\textbf{c}))+\sum_{i=0}^{k-1} b_i\eta^{k-i}(\pi^{(n-m)/2}(\textbf{c})))
$$
Hence, $(\overline{\zeta}^{-1}(\textbf{g}))^r\in \overline{\zeta}^{-1}(C_{\eta}) $.

Then $\overline{\zeta}^{-1}(C_{\eta})$ is a reversible DNA code. By the same argument we can get the proofs of the remainder situations, where $C_{\eta}=\langle S_t \rangle_ \eta$, $C_{\eta}=\langle E_t \rangle_ \eta$, $C_{\eta}=\langle E_t^{(1)} \rangle_ \eta$ and $C_{\eta}=\langle E_t^{(2)} \rangle_ \eta$.\\

\textbf{Example 4.11} Let $\textbf{c}=(1,\alpha^2,1,\alpha^2,\alpha^2,\alpha,1,1,\alpha,\alpha^2,\alpha^2 )\in F_4^{11}$ . Obviously, $c(x)$ is a DNA-$3$-quasi-reciprocal polynomial over $F_{4}$. For $t=0$, we can get $\mathcal{C}={\langle}E_0^{(2)}{\rangle}={\langle}{\pi}^0(\textbf{c}),{\pi}^{-3}(\textbf{c}),{\pi}^4(\textbf{c}){\rangle}$, and the generator matrix of code $\mathcal{C}$ is
$$\textbf{G}=\left(
\begin{array}{ccccccccccc}
1 & {\alpha}^2 & 1 & {\alpha}^2 & {\alpha}^2 & {\alpha} & 1 & 1 & {\alpha} & {\alpha}^2 & {\alpha}^2 \\
{\alpha}^2 & {\alpha}^2 & {\alpha} & 1 & 1 & {\alpha} & {\alpha}^2 & {\alpha}^2 & 1 & {\alpha}^2 & 1 \\
1 & {\alpha} & {\alpha}^2 & {\alpha}^2 & 1 & {\alpha}^2 & 1 & {\alpha}^2 & {\alpha}^2 & {\alpha} & 1  \\
\end{array}
\right)
$$
Then we get a optimal code $[11,3,7]$ added property of being reversible.\\

\textbf{Example 4.12} Let $\emph{B}^{\emph{DNA}}_3=(TTT,AAA,TTT,TAA,ATA,ATA,AAT)$, $\textbf{c}={\overline{\zeta}}(\emph{B}^{\emph{DNA}}_3) =(1,0,1,\gamma,\gamma^2,\gamma^2,1+\gamma+\gamma^2)\in F_{64}^7$.  Clearly, $c(x)$ is a DNA-$3$-quasi-reciprocal polynomial over $F_{64}$,
For $t=0$, we can get a map-code $\mathcal{C}_{\eta}={\langle}\pi^0(\textbf{c}),\pi^{-3}(\textbf{c}){\rangle}_\eta$.
Then the codewords of $\mathcal{C}_\eta$ are generated as follows\\
$\mathcal{C}_\eta={\{}a_1\pi^{-3}(\textbf{c})+a_2\eta(\pi^{-3}(\textbf{c}))+a_3\eta^2(\pi^{-3}(\textbf{c}))+b_1(\pi^{0}(\textbf{c}))+b_2\eta(\pi^{0}(\textbf{c}))+ b_3\eta^2(\pi^{0}(\textbf{c})):$ \\$a_1, a_2, a_3, b_1, b_2, b_3\in F_{4}{\}},$
 and the corresponding DNA code is a reversible DNA code.

For $b_2=\alpha$, $b_1=b_3=a_1=a_2=a_3=0$,
we obtain the codeword
$${\textbf{c}_1}=(\alpha,0,\alpha,\alpha+\alpha \gamma+\alpha \gamma^2,\alpha{\gamma},\alpha \gamma,\alpha{\gamma}^{2}).$$
The corresponding DNA codeword is
$$D_{\textbf{c}_1}=\overline{\zeta}^{-1}(\textbf{c}_1)=(CCCAAACCCAACCAACAAACA).$$

For the reverse of the DNA codeword, we can take ${{a}_1}=a_2=b_1=b_2=b_3=0$, ${a}_3=\alpha$ by Lemma 4.8. Then the codeword is
$$\textbf{c}_2=(\alpha{\gamma}^{2},\alpha+\alpha \gamma+\alpha{\gamma}^{2},\alpha+\alpha \gamma+\alpha{\gamma}^{2},\alpha \gamma,\alpha,0,\alpha).$$
The corresponding DNA codeword is
$$D_{\textbf{c}_2}=\overline{\zeta}^{-1}(\textbf{c}_2)=(ACAAACAACCAACCCAAACCC).$$
So we can find that, $({D_{\textbf{c}_1}})^r=D_{\textbf{c}_2}.$

If we take $\textbf{c}_3=\pi^0(\textbf{c})+\alpha\eta(\pi^{-3}(\textbf{c}))=(\alpha^2+ \alpha \gamma+\alpha \gamma^{2},\alpha \gamma,1+\alpha \gamma,\gamma+\alpha\gamma^2,\alpha+\gamma^2,\gamma^{2},\alpha^2+\gamma +\gamma^2)$, we can get the corresponding DNA codeword
$$D_{\textbf{c}_3}=(TTGCAAGTTTCACGCATACCG).$$
For the reverse of the DNA codeword, we can take $a_2=b_1=b_2=b_3=0$, ${a}_1=1,{a}_3=\alpha$ by Lemma 4.8. Then the codeword is
$$\textbf{c}_4={\pi}^{-3}(\textbf{c})+{\alpha}{\eta}^2({\pi}^0(\textbf{c}))=(\alpha+{\gamma},{\gamma}^{2},\alpha+{\gamma}^2,1+\gamma+\alpha^2{\gamma}^2,
\alpha^2+\alpha\gamma+\alpha{\gamma}^2,\alpha+\alpha \gamma+ \alpha \gamma^2,1+\alpha{\gamma}).$$\\
the DNA codeword is $D_{\textbf{c}_4}=(GCCATACGCACTTTGAACGTT).$
So we can find that $({D_{\textbf{c}_3}})^r=D_{\textbf{c}_4}.$\\

\textbf{Theorem 4.13} Let $c(x)\in F_{4^k}[x]/\langle x^n-1\rangle$ be a DNA-$m$-quasi-reciprocal polynomial and $r(x)=1+x+x^2+\cdots+x^{n-1}$. Let the set $S^c={\{}S_t,\textbf{r}{\}}$, $E^c={\{}E_t,\textbf{r}{\}}$, $E^{(1)c}={\{}E_t^{(1)},\textbf{r}{\}}$, $E^{(2)c}={\{}E_t^{(2)},\textbf{r}{\}}$ and $E^{(3)c}={\{}E_t^{(3)},\textbf{r}{\}}$,  then $C_{\eta}={\langle}S^c{\rangle}_{\eta}$, $C_{\eta}=\langle E^c\rangle_{\eta}$, $C_{\eta}=\langle E^{(1)c}\rangle_{\eta}$, $C_{\eta}=\langle E^{(2)c}\rangle_{\eta}$ or $C_{\eta}=\langle E^{(3)c}\rangle_{\eta}$ is a map-code , where the set $S_t$, $E_t$, $E_t^{(1)}$, $E_t^{(2)}$ or $E_t^{(3)}$ comes from theorem 3.4. Then $\overline{\zeta}^{-1}(C_{\eta})$ is a reversible complement DNA code.\\

\textbf{Proof}: ${\langle}S^c{\rangle}$, ${\langle}E^{c}{\rangle}$, ${\langle}E^{(1)c}{\rangle}$, ${\langle}E^{(2)c}{\rangle}$ or ${\langle}E^{(3)c}{\rangle}$ is a reversible code by Lemma 3.3, then  $C_{\eta}$ is a map-code. The proof of $\overline{\zeta}^{-1}(C_{\eta})$ is a reversible DNA code is similar to the proof in Theorem 4.10. Furthermore, $\textbf{r} \in C_{\eta}$ and $\overline{\zeta}(x^c)=\overline{\zeta}(x)+1$ for $x\in D^k$, then $\overline{\zeta}^{-1}(z+1)=(\overline{\zeta}^{-1}(z))^c$, for $z \in F_{4^k}$, then $\overline{\zeta}^{-1}(C_{\eta})$ is a reversible complement DNA code.\\

\textbf{Example 4.14} As Example 4.12, we take $\emph{B}^{\emph{DNA}}_3=(TTT,AAA,TTT,TAA,ATA,ATA,AAT)$. And we take $\mathcal{C}_{\eta}={\langle}\pi^0(\textbf{c}),\pi^{-3}(\textbf{c}),\textbf{r}{\rangle}_\eta$. Then $\overline{\zeta}^{-1}(C_{\eta})$ is a reversible complement DNA code. Then the codewords of $\mathcal{C}_\eta$ are generated as follows\\
$\mathcal{C}_\eta={\{}a_1\pi^{-3}(\textbf{c})+a_2\eta(\pi^{-3}(\textbf{c}))+a_3\eta^2(\pi^{-3}(\textbf{c}))+b_1(\pi^{0}(\textbf{c}))+b_2\eta(\pi^{0}(\textbf{c}))+ b_3\eta^2(\pi^{0}(\textbf{c}))+b_4 \textbf{r}:$ \\$a_1, a_2, a_3, b_1, b_2, b_3, b_4\in F_{4}{\}}$.

For $b_2=\alpha$, $b_4=\alpha^2$, $b_1=b_3=a_1=a_2=a_3=0$,
we obtain the codeword
$${\textbf{c}_5}=(1,\alpha^2,1,1+\alpha{\gamma}+\alpha\gamma^2,\alpha^2+\alpha{\gamma},\alpha^2+\alpha\gamma,\alpha^2+\alpha{\gamma}^2).$$
The corresponding DNA codeword is
$$D_{\textbf{c}_5}=\overline{\zeta}^{-1}(\textbf{c}_5)=(TTTGGGTTTGGTTGGTGGGTG).$$

For the reverse of the DNA codeword, we can take ${{a}_1}=a_2=b_1=b_2=b_3=0$, ${a}_3=\alpha$,  ${b}_4=\alpha^2$ by Lemma 4.8. Then the codeword is
$$\textbf{c}_6=(\alpha^2+\alpha{\gamma}^2,1+\alpha\gamma+\alpha{\gamma}^2,1+\alpha\gamma+\alpha{\gamma}^2,\alpha^2+\alpha{\gamma},1,\alpha^2,1).$$
The corresponding DNA codeword is
$$D_{\textbf{c}_6}=\overline{\zeta}^{-1}(\textbf{c}_6)=(GTGGGTGGTTGGTTTGGGTTT).$$
So we can find that, $({D_{\textbf{c}_5}})^r=D_{\textbf{c}_6}.$

The reversible complement codeword of $\textbf{c}_6$ is $\textbf{c}_7=\textbf{c}_6 + \textbf{r}$.
$$D_{\textbf{c}_7}=\overline{\zeta}^{-1}(\textbf{c}_7)=(CACCCACCAACCAAACCCAAA).$$

\begin{table}[htbp]
	\begin{center}
		\caption{\small $[11,3,7]$ reversible DNA code according to Example $4.11$.}
		\begin{tabular}{cccc}
			\hline
			$AAAAAAAAAAA$ & $TGTGGCTTCGG$ & $CTCTTGCCGTT$ & $GCGCCTGGTCC$ \\
			$TCGGTGTGGCT$ & $ATCACTACTTC$ & $GGTCAAGTAGA$ & $CAATGCCACAG$ \\
			$CGTTCTCTTGC$ & $GAACTGGAGAT$ & $ACGAGCAGCCG$ & $TTCGAATCATA$ \\
			$GTCCGCGCCTG$ & $CCGTAACGACA$ & $TAAGCTTATAC$ & $AGTATGATGGT$ \\
			$GGCTTCGGTGT$ & $CAGCCACCGAC$ & $TCAAATTTCCA$ & $ATTGGGAAATG$ \\
			$CTTCATCACTA$ & $GCATGGGTACG$ & $AAGGTCACTAT$ & $TGCACATGGGC$ \\
			$TAGAGGTCAAG$ & $AGCGATAGCGA$ & $GTTTCAGAGTC$ & $CCACTCCTTCT$ \\
			$ACAGCAATGCC$ & $TTTATCTATTT$ & $CGCCGGCGAGG$ & $GAGTATGCCAA$ \\
			$TTGCCGTTCTC$ & $ACCTTTAAACT$ & $GATGGAGGTAG$ & $CGAAACCCGGA$ \\
			$AGATGAACTGG$ & $TATCACTGGAA$ & $CCCACGCACCC$ & $GTGGTTGTATT$ \\
			$GCCGACGAGCA$ & $CTGAGACTTTG$ & $TGACTTTCAGT$ & $AATTCGAGCAC$ \\
			$CATATTCGAAT$ & $GGAGCGGCCGC$ & $ATGTACATGTA$ & $TCCCGATATCG$ \\
			$CCTGGTCCGCG$ & $GTAAAGGGTTA$ & $AGGCCCAAAGC$ & $TACTTATTCAT$ \\
			$GACACCGTAAC$ & $CGGGTACACGT$ & $TTATGTTGGTG$ & $ACTCAGACTCA$ \\
			$ATACTAAGCTT$ & $TCTTCCTCACC$ & $CACGAGCTTAA$ & $GGGAGTGAGGG$ \\
			$TGGTAGTATGA$ & $AACCGTATGAG$ & $GCTATAGCCCT$ & $CTAGCCCGATC$ \\
			\hline
		\end{tabular}
	\end{center}
\end{table}

\dse{5~~Conclusion}

We have described a novel polynomial, $m$-quasi-reciprocal polynomial, to construct reversible codes over finite fields. This polynomial is generalized from coterm polynomial in [15]. This novel approach, constructing reversible codes, is proved to be very advantageous which we can determine the length $n$, and we may change the dimensions of reversible codes. We find some reversible codes over $F_q$ have the same parameters with the best known in the Datebase or reach the Griesmer bound. We give a  map from DNA $k$-bases to the elements of $F_{4^k}$. Then we construct reversible DNA codes over $F_{4^k}$ by DNA-$m$-quasi-reciprocal polynomials.

\end{document}